\documentclass{scrartcl}

\KOMAoptions{paper=a4,paper=portrait,fontsize=10pt}
\usepackage[english]{babel}
\usepackage[ansinew]{inputenc}
\usepackage[T1]{fontenc}
\usepackage{helvet}
\usepackage[a4paper,portrait]{geometry}
\let\oldtwocolumn\twocolumn

\newif\iftwocolumn\twocolumntrue
\def\onecolumn{\twocolumnfalse}
\def\twocolumn{\twocolumntrue}
\usepackage{indentfirst}
\newlength{\OneColumnWidth}
\newlength{\TwoColumnWidth}
\makeatletter
\renewcommand\section{\scr@startsection{section}{1}{\z@}{-3.5ex \@plus -1ex \@minus -.2ex}{2.3ex \@plus.2ex}{\normalfont\bfseries}}
\renewcommand\subsection{\scr@startsection{subsection}{2}{\z@}{-3.5ex \@plus -1ex \@minus -.2ex}{2.3ex \@plus.2ex}{\normalfont\bfseries}}
\makeatother

\usepackage[
	colorlinks=true,
	urlcolor=blue, 
	filecolor=blue, 
	linkcolor=blue, 
	citecolor=blue, 
]{hyperref}
\usepackage{cite} 

\usepackage[singlelinecheck=true,font=small,labelfont=bf,format=plain]{caption} 

\let\oldhline\hline
\renewcommand{\hline}{\oldhline\rule{0pt}{12pt}}
\usepackage{enumitem}
\setlist{nosep}
\setenumerate[1]{label=(\arabic*)}
\setenumerate[2]{label=(\Alph*)}

\usepackage{ifthen}
\usepackage{forloop}
\usepackage{etoolbox}
\usepackage{graphicx}
\usepackage[fleqn]{amsmath}

\renewcommand{\title}[1]{\def\inserttitle{#1}}
\newcommand{\email}[1]{\def\insertemail{#1}}
\renewcommand{\abstract}[1]{\def\insertabstract{#1}}
\def\insertjournal{}
\def\insertdoi{}
\def\insertarxiv{}
\newcommand{\journal}[3][accepted]{
	\def\tmpa{#1}\def\tmpb{submitted}\ifx\tmpa\tmpb\def\journalpre{Submitted to: }\else\def\tmpb{accepted}\ifx\tmpa\tmpb\def\journalpre{Accepted for publication in: }\else\def\tmpb{prepared}\ifx\tmpa\tmpb\def\journalpre{Prepared for submission to: }\else\def\journalpre{}\fi\fi\fi
	\if\relax\detokenize{#3}\relax\def\insertjournal{\journalpre#2}\else\def\insertjournal{\journalpre\href{#3}{#2}}\fi
}
\newcommand{\doi}[1]{\if\relax\detokenize{#1}\relax\def\insertdoi{}\else\def\insertdoi{DOI: \href{http://dx.doi.org/#1}{#1}}\fi}
\newcommand{\arxiv}[2]{\if\relax\detokenize{#2}\relax\def\insertarxiv{}\else\def\insertarxiv{arXiv: \href{https://arxiv.org/abs/#1}{#1 [#2]}}\fi}
\newcounter{authors}
\newcounter{addresses}
\newcounter{keywords}
\newcommand{\addauthor}[2]{\csdef{author\arabic{authors}}{#1}\csdef{authoraddress\arabic{authors}}{#2}\stepcounter{authors}}
\newcommand{\addaddress}[1]{\csdef{address\arabic{addresses}}{#1}\stepcounter{addresses}}
\newcommand{\addkeyword}[1]{\csdef{keyword\arabic{keywords}}{#1}\stepcounter{keywords}}
\newcommand{\maketitlesub}{
	\newcounter{i}
	\newcounter{j}
	\noindent\textbf{%
		\Large\inserttitle\\[1em]
		\large\csuse{author0}$^{\csuse{authoraddress0}}$%
		\forloop{i}{1}{\value{i} < \value{authors}}{%
			, \csuse{author\arabic{i}}$^{\csuse{authoraddress\arabic{i}}}$%
		}
	}\\[1em]
	\normalsize
	\setcounter{j}{0}
	\forloop{i}{0}{\value{i} < \value{addresses}}{%
		\stepcounter{j}
		\ifnum\value{addresses}>1$^{\arabic{j}}$\fi\,\csuse{address\arabic{i}}
		\ifthenelse{\value{j}<\value{addresses}}{\\}{}
	}
	\ifx\insertemail\empty\\[1em]\else\\[0.5em]E-mail address: \insertemail\\[1em]\fi
	\textbf{Abstract:} \insertabstract
	\ifthenelse{\value{keywords}=0}{}{
		\\[1em]
		Keywords: \csuse{keyword0}%
			\forloop{i}{1}{\value{i} < \value{keywords}}{%
				; \csuse{keyword\arabic{i}}%
			}
	}
}
\renewcommand{\maketitle}{\iftwocolumn\oldtwocolumn[\maketitlesub\vspace{1.5em}]\else\maketitlesub\fi}

\usepackage[headsepline]{scrlayer-scrpage}
\clearpairofpagestyles
\ihead{%
	\ifx\insertarxiv\empty%
			\ifx\insertjournal\empty\else\textnormal\insertjournal\fi%
	\else%
		\ifx\insertdoi\empty%
			\ifx\insertjournal\empty\else\textnormal\insertjournal\fi%
		\else%
			\ifx\insertjournal\empty\else\textnormal\insertjournal\fi%
			\ifx\textnormal\empty\else\linebreak\textnormal\insertdoi\fi%
		\fi%
	\fi%
}
\ohead{%
	\ifx\insertarxiv\empty%
		\ifx\textnormal\empty\else\textnormal\insertdoi\fi%
	\else%
		\ifx\insertdoi\empty%
			\ifx\insertarxiv\empty\else\textnormal\insertarxiv\fi%
		\else%
			\ifx\insertarxiv\empty\else\linebreak\textnormal\insertarxiv\fi%
		\fi%
	\fi%
}
\ifoot{}
\ofoot{}

\usepackage[hang]{footmisc}
\let\oldthebibliography\thebibliography

\renewcommand\thebibliography[1]{
	\small
	\oldthebibliography{#1}
	\setlength{\parskip}{0pt}
	\setlength{\itemsep}{0pt plus 0.3ex}
}
\newcommand{\bstindent}{99}
\newcommand{\bstaddress}{}
\newcommand{\bstauthor}{}

\newcommand{\bstjournal}{}

\newcommand{\bstpublisher}{}

\newcommand{\bsttitle}{}
\newcommand{\bstvolume}{}
\newcommand{\bstyear}{}
\newcommand{\bbland}{and}

\newcommand{\bblnov}{November}

\usepackage{empheq}
\usepackage{upgreek}

\usepackage{dblfloatfix} 
\allowdisplaybreaks 

\renewcommand{\hbar}{\mathchar'26\mkern-9mu \mathrm{h}}

\newcommand{\hamilton}{\mathcal{H}}
\newcommand{\overlap}{\mathcal{O}}
\newcommand{\coupling}{\tau}

\newcommand{\overlapcoupling}{\nu}
\newcommand{\green}{\mathcal{G}}

\newcommand{\transmission}{\mathcal{T}}
\newcommand{\dos}{\mathcal{D}}

\newcommand{\imag}{\text{i}}

\renewcommand{\Im}{\text{Im}~}

\DeclareMathOperator{\Tr}{Tr}
\newcommand{\mydots}{\rotatebox{-20}{$\cdots$}}

\renewcommand{\hamilton}{\mathcal{H}}
\renewcommand{\coupling}{\tau}
\renewcommand{\overlap}{\mathcal{S}}
\renewcommand{\overlapcoupling}{\mathcal{S}}
\renewcommand{\green}{\mathcal{G}}
\renewcommand{\transmission}{\mathcal{T}}
\renewcommand{\dos}{\mathcal{D}}
\renewcommand{\imag}{\mathrm{i}}

\renewcommand{\Im}{\mathrm{Im}}
\renewcommand{\mydots}{\rotatebox{-20}{$\cdots$}}

\begin{document}

\onecolumn 

\title{An improved Green's function algorithm applied to quantum transport in carbon nanotubes}

\addauthor{Fabian Teichert}{1,3,4}
\addauthor{Andreas Zienert}{2}
\addauthor{J\"org Schuster}{3,4}
\addauthor{Michael Schreiber}{1,4}

\addaddress{Institute of Physics, Chemnitz University of Technology, 09107 Chemnitz, Germany}
\addaddress{Center for Microtechnologies, Chemnitz University of Technology, 09107 Chemnitz, Germany}
\addaddress{Fraunhofer Institute for Electronic Nano Systems (ENAS), 09126 Chemnitz, Germany}
\addaddress{Dresden Center for Computational Materials Science (DCMS), TU Dresden, 01062 Dresden, Germany}

\email{fabian.teichert@physik.tu-chemnitz.de}

\abstract{
The renormalization-decimation algorithm (RDA) of L\'opez Sancho et al. is used in quantum transport theory to calculate bulk and surface Green's functions.
We derive an improved version of the RDA for the case of very long quasi one-dimensional unit cells (in transport direction).
This covers not only long unit cells but also supercell-like calculations for structures with disorder or defects.
In such large systems, short-range interactions lead to sparse real-space Hamiltonian matrices.
We show how this and a corresponding subdivision of the unit cell in combination with the decimation technique can be used to reduce the calculation time.
Within the resulting algorithm, separate RDA calculations of much smaller effective Hamiltonian matrices must be done for each Green's function, which enables the treatment of systems too large for the common RDA.
Finally, we discuss the performance properties of our improved algorithm as well as some exemplary results for chiral carbon nanotubes.
}

\addkeyword{renormalization-decimation algorithm (RDA)}
\addkeyword{electronic transport}
\addkeyword{carbon nanotube (CNT)}

\journal[]{Computational Materials Science 169 (2019) 109014}{https://doi.org/10.1016/j.commatsci.2019.05.012} 
\doi{10.1016/j.commatsci.2019.05.012} 
\arxiv{1806.02039}{physics.comp-ph} 

\maketitle

\section{Introduction}

Over the last decades simulation techniques became an important tool for determining material properties.
The possibilities of quantum simulation enhanced a lot in the past.
The rapid progress in micro- and nanoelectronics results in a further miniaturization and new and better devices.
This leads to a convergence of both, microelectronics and quantum mechanics, whereby devices get accessible with quantum simulation techniques.
Simulations are widely used for the theoretical investigation of materials, but also for the discovery of completely new materials.
Today's computer power enables high throughput material development, e.g. by scanning through ternary alloys of different element combinations all over the periodic system~\cite{NatureMaterials.12.191}.
The treatment of huge systems in the mesoscopic range is possible, too, enabling investigations of long-range effects.
A third perspective is to enhance the accuracy of material and device modeling by including more and more effects like electron-electron interaction, electron-phonon interaction, spin-orbit coupling, etc.

In the present study, we treat the topic of quantum transport simulations~\cite{Datta2005,PhysRevB.31.6207}, which is based on the theoretical framework of Green's functions.
They are widely used for calculating, e.g., the conductance of novel molecular conductors~\cite{JPhysChemLett.4.809}, the performance of new field effect transistors (FET)~\cite{Nanoscale.8.10240}, and the functionality of sensors~\cite{JPhysChemC.112.13442}.
The present work focuses on the treatment of systems in the mesoscopic range and improves quantum transport algorithms to reduce the computation time.
Density functional theory (DFT)~\cite{BrazJPhys.36.1318,PhysRevB.65.165401}, as the underlying electron structure theory, can handle up to some thousands of atoms, but can still not reach the region of hundreds of thousands of atoms.
For this purpose, tight-binding (TB) models are frequently used.
Whereas simple distant-independent nearest-neighbor models often only provide some qualitative understanding, the density-functional-based tight-binding (DFTB) model~\cite{PhysRevB.51.12947,IntJQuantumChem.58.185} combines the speed of TB calculations with DFT-like accuracy.
Many DFTB parameter sets have been published, describing different systems correctly, organic molecules on the one side, metals on the other side, but also mixed systems like devices with organic molecules connected to metallic electrodes.
Such DFTB models can be used as a basis for calculating the electronic structure and the electronic transport of mesoscopic systems.

Examples suitable for using DFTB are systems with random disorder~\cite{PhysRevLett.99.076803,JPhysCondMat.20.294214,JPhysCondMat.20.304211,NJPhys.16.123026,JPhysCondMat.26.045303,JComputElectron.17.521,JPhysCommun.2.105012}.
They are often treated with recursive or iterative techniques.
The linearly scaling recursive Green's function formalism (RGF)~\cite{JPhysCSolidStatePhys.5.2845,CompPhysCommun.20.11,JPhysCSolidStatePhys.14.235,ZPhysBCondMat.59.385} and the logarithmically scaling renormalization-decimation algorithm (RDA) of L\'opez Sancho et al.~\cite{JPhysFMetPhys.14.1205,JPhysFMetPhys.15.851} can be used to calculate the electron transmission in such a system.
Both algorithms are based on decimation, i.e., subdividing the total system into pieces.
Long unit cells (UCs) can be further subdivided.
The RGF can be easily customized for this case.
But as the RDA requires periodicity, it has to be rewritten to get an improvement here as well.
In the following, we address this problem and show how the RDA can be adapted to quasi one-dimensional systems with long UCs to reduce the calculation time.
This further enables quantum transport calculations for systems too large to treat with the common RDA within acceptable time.

First, we derive the equations of the improved RDA to calculate the surface and bulk Green's functions, the transmission, and the bulk density of states.
Second, we consider the scaling of the computational complexity.
For this, we perform computations for chiral carbon nanotubes (CNTs) with different UC lengths to verify the complexity dependencies for one example.
Finally, we depict some examplary results for defective (10,1)- and (10,9)-CNTs, which can be used in future microelectronic devices.

\section{Electronic transport}

\begin{figure}
	\centering
	\includegraphics{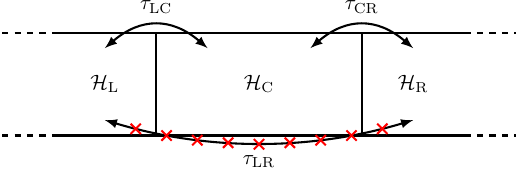}
	\caption[Scheme of a device system]{(Color online.) Device scheme~\cite{NJPhys.16.123026}. A system with infinite length is divided into a finite central region C and two semi-infinite electrodes L (left) and R (right), connected to C. C contains an arbitrary scattering region. $\hamilton_\mathrm{L/C/R}$ are the Hamiltonian matrices of the corresponding parts L/C/R. $\coupling_\mathrm{LC/CR/LR}$ are the coupling matrices connecting two of these parts, but $\coupling_\mathrm{LR}$ is assumed to be negligible.}\label{JCP2:fig:Device}
\end{figure}

In this section we give a brief overview about the common equilibrium quantum transport theory~\cite{Datta2005}, which tells us how to calculate electronic properties like density of states, electron density, transmission, and conductance.
Let us consider the device system shown in figure \ref{JCP2:fig:Device}.
This is an infinite system with a scattering region at the center (C), which is finite in transport direction, and semi-infinite regions to the left (L) and the right (R) of C, the electrodes.
The whole quasi one-dimensional system is not periodic in the plane perpendicular to the transport direction.
The electrodes act as the reservoirs of an open system and provide and absorb electrons, which are transmitted through the system and scattered within the scattering region C.
For such a device, the Schr\"odinger equation within a real-space non-orthogonal basis reads
\begin{equation}
	\hamilton_\mathrm{tot}\varPsi = E\overlap_\mathrm{tot}\varPsi \quad\text{with}\quad \hamilton_\mathrm{tot} = \begin{pmatrix}
		\hamilton_\mathrm{L} & \coupling_\mathrm{LC} & ~ \\
		\coupling_\mathrm{CL} & \hamilton_\mathrm{C} & \coupling_\mathrm{CR} \\
		~ & \coupling_\mathrm{RC} & \hamilton_\mathrm{R}
	\end{pmatrix} \quad ,\quad \overlap_\mathrm{tot} = \begin{pmatrix}
		\overlap_\mathrm{L} & \overlap_\mathrm{LC} & ~ \\
		\overlap_\mathrm{CL} & \overlap_\mathrm{C} & \overlap_\mathrm{CR} \\
		~ & \overlap_\mathrm{RC} & \overlap_\mathrm{R}
	\end{pmatrix} \quad .\label{JCP2:eqn:Schroedinger}
\end{equation}%
$\hamilton_\mathrm{L/C/R}$ are the Hamiltonian matrices of the corresponding regions.
$\coupling_\mathrm{LC/CL/RC/CR}$ are the Hamiltonian coupling matrices between two of these regions.
If the central region is large enough, i.e. larger than the maximum interaction distance, the coupling between L and R can be neglected, $\coupling_\mathrm{LR/RL}=0$.
$\overlap_\mathrm{L/C/R}$ are the corresponding overlap matrices.
$\overlap_\mathrm{LC/CL/RC/CR}$ are the overlap matrices for the coupling of two of these regions.
For fixed $E$, we can get rid of the $\overlap$~terms by hiding them in the $\hamilton$ and $\coupling$ terms with the substitution
\begin{equation}
	\hamilton := \hamilton - E(\overlap-\mathcal{I}) \quad,\quad \coupling := \coupling - E\overlapcoupling \quad.
\end{equation}%
$\mathcal{I}$ is the identity matrix of appropriate dimension.
After this, (\ref{JCP2:eqn:Schroedinger}) transforms into an orthonormal-like equation with the new $\hamilton$ and $\coupling$, which will be used within all subsequent explanations.

Equation (\ref{JCP2:eqn:Schroedinger}) has infinite matrix dimension, but can be reduced to a problem of finite dimension by calculating the perturbed Green's function of the central region
\begin{equation}
	\green_\mathrm{C} = \left[(E+\imag\eta)\mathcal{I}-\hamilton_\mathrm{C}-\varSigma_\mathrm{L}-\varSigma_\mathrm{R}\right]^{-1} \quad.
\end{equation}%
Here, the difficulty of treating an infinite system is shifted to the calculation of the self-energies $\varSigma_\mathrm{L}=\coupling_\mathrm{CL}g_\mathrm{L}\coupling_\mathrm{LC}$ and $\varSigma_\mathrm{R}=\coupling_\mathrm{CR}g_\mathrm{R}\coupling_\mathrm{RC}$, which lead to energetic shifts of the electronic states within C due to the coupling to the left and right electrode.
The surface Green's functions of the left electrode $g_\mathrm{L}$ and the right electrode $g_\mathrm{R}$ can be calculated iteratively with the RDA shown in section \ref{JCP2:sec:RDA}.

With the use of the Green's function the transmission spectrum can be determined as
\begin{equation}
	\transmission(E) = \Tr\left(\varGamma_\mathrm{R}\green_\mathrm{C}\varGamma_\mathrm{L}\green_\mathrm{C}^\dagger\right) \quad.
\end{equation}%
The broadening matrices $\varGamma_\mathrm{L/R} = \imag(\varSigma_\mathrm{L/R}-\varSigma_\mathrm{L/R}^\dagger)$ lead to a broadening of the electronic states within C due to the coupling to the electrodes.
The conductance of the total system can be calculated using the Landauer-B\"uttiker formalism~\cite{PhysRevB.31.6207}
\begin{equation}
	G = -\mathrm{G}_0\int\limits_{-\infty}^\infty\transmission(E)\frac{\mathrm{d}f}{\mathrm{d}E}\,\mathrm{d}E \qquad\mathrm{with}\qquad \mathrm{G}_0=\frac{2\mathrm{e}^2}{\mathrm{h}} \qquad\mathrm{and}\qquad f(E)=\frac{1}{1+\mathrm{exp}\left(\frac{E-E_\mathrm{F}}{k_\mathrm{B}T}\right)} \quad.
\end{equation}%
$E_\mathrm{F}$ is the Fermi energy.

\section{Decimation technique and renormalization-decimation algorithm}\label{JCP2:sec:RDA}

The RDA was derived by L\'opez Sancho et al.~\cite{JPhysFMetPhys.14.1205,JPhysFMetPhys.15.851}.
It is an iterative algorithm to calculate the bulk and surface Green's functions ($\green_\mathrm{B/L/R}$) of periodic systems, e.g. the electrodes of a device described above.
Note, that in the following, subscripting is done from an electrode's point of view instead of the device's point of view.
The surface Green's function of the left electrode $g_\mathrm{L}$ equals the Green's function at the right surface $\green_\mathrm{R}$ of the semi-infinite system L.
The real-space Hamiltonian matrix of the electrode is periodic, because the electrode is periodic.
Assuming short-range interaction it is also block-wise tridiagonal.
Considering a large, but finite system, the electrode Hamiltonian matrix and the corresponding Green's matrix read
\begin{equation}
	\hamilton_\mathrm{el} = \begin{pmatrix}
		\hamilton & \coupling & ~ & ~ & ~ \\
		\coupling^\dagger & \hamilton & \coupling & ~ & ~ \\
		~ & \coupling^\dagger & \mydots & \mydots & ~ \\
		~ & ~ & \mydots & \hamilton & \coupling \\
		~ & ~ & ~ & \coupling^\dagger & \hamilton
	\end{pmatrix} \quad\mathrm{and}\quad \green_\mathrm{el} = \begin{pmatrix}
		\green_\mathrm{L} & \cdots & ~ & ~ & ~ \\
		\smash{\vdots} & \mydots & \mydots & ~ & ~ \\
		~ & \mydots & \green_\mathrm{B} & \mydots & ~ \\
		~ & ~ & \mydots & \mydots & \smash{\vdots} \\
		~ & ~ & ~ & \cdots & \green_\mathrm{R}
	\end{pmatrix} \quad.\label{JCP2:eqn:Helectrode}
\end{equation}%
$\hamilton$ is the Hamiltonian matrix of one UC.
$\coupling$ is the Hamiltonian coupling matrix between adjacent UCs.
$\green_\mathrm{B}$ is the bulk Green's matrix corresponding to the bulk-like periodic region far away from the finite ends.
$\green_\mathrm{L}$ is the surface Green's matrix of the left surface and $\green_\mathrm{R}$ the one of the right surface.
As only the three mentioned Green's matrix blocks are of interest, the decimation technique can be used to reduce the inversion problem of the total system to an inversion problem of a smaller effective system containing the relevant parts and ignoring the rest.

\begin{figure}
	\centering
	\includegraphics{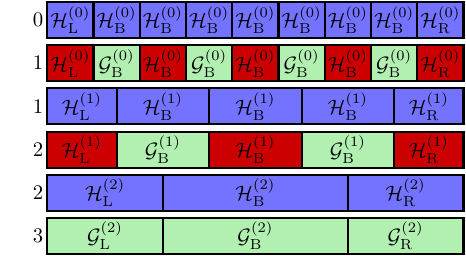}
	\caption[Sketch of the RDA]{(Color online.) Sketch of the RDA for an electrode with 9 UCs. The numbers to the left of each row indicate the iteration step $i$ in (\ref{JCP2:eqn:RDA}). Blue blocks represent the Hamiltonian matrices. Green blocks signify the Green's matrices of the cells decimated within the respective iteration step. Red blocks denote Hamiltonian matrices which get a correction due to the coupling to decimated adjacent cells.}\label{JCP2:fig:RDA}
\end{figure}

At first, let us consider a given 3$\times$3 matrix $A$ and its unknown inverse $B$.
Let us further assume that only the upper left and lower right entries $B_{11}$ and $B_{33}$ of $B$ are of interest.
For that, the 3$\times$3 inversion problem
\begin{equation}
	\begin{pmatrix}
		A_{11} & A_{12} & A_{13} \\
		A_{21} & A_{22} & A_{23} \\
		A_{31} & A_{32} & A_{33}
	\end{pmatrix}\begin{pmatrix}
		B_{11} & B_{12} & B_{13} \\
		B_{21} & B_{22} & B_{23} \\
		B_{31} & B_{32} & B_{33}
	\end{pmatrix} = \begin{pmatrix}
		1 & 0 & 0 \\
		0 & 1 & 0 \\
		0 & 0 & 1
	\end{pmatrix}\label{JCP2:eqn:RDA:inversion1}
\end{equation}%
can be reduced to the 2$\times$2 inversion problem
\begin{equation}
	A_\mathrm{eff}\begin{pmatrix}
		B_{11} & B_{13} \\
		B_{31} & B_{33}
	\end{pmatrix} = \begin{pmatrix}
		1 & 0 \\
		0 & 1
	\end{pmatrix} \quad.\label{JCP2:eqn:RDA:inversion2}
\end{equation}%
Comparing (\ref{JCP2:eqn:RDA:inversion1}) and (\ref{JCP2:eqn:RDA:inversion2}), the effective matrix $A_\mathrm{eff}$ is given by
\begin{equation}
	A_\mathrm{eff} = \begin{pmatrix}
		A_{11} & A_{13} \\
		A_{31} & A_{33}
	\end{pmatrix} + \begin{pmatrix}
		A_{12} \\
		A_{32}
	\end{pmatrix} A_{22}^{-1} \begin{pmatrix}
		A_{21} & A_{23}
	\end{pmatrix} = \begin{pmatrix}
		A_{11} + A_{12}A_{22}^{-1}A_{21} & A_{13} + A_{12}A_{22}^{-1}A_{23} \\
		A_{31} + A_{32}A_{22}^{-1}A_{21} & A_{33} + A_{32}A_{22}^{-1}A_{23}
	\end{pmatrix} \quad.\label{JCP2:eqn:decimation:Aeff}
\end{equation}%
Applying the latter to the electrode Hamiltonian and Green's matrix, an iterative algorithm can be derived, where $A_{13}=A_{31}=0$ because of the absence of coupling beyond adjacent cells.
Utilizing a divide-and-conquer strategy the RDA can be obtained.
A graphical visualization of the RDA is shown in figure \ref{JCP2:fig:RDA}.
An electrode with $2^k+1$ UCs is considered and every second cell is decimated in each step until only three cells are left.
They correspond to the left/bulk/right Green's matrix.
This leads to the following iterative set of equations:
\begin{subequations}
	\label{JCP2:eqn:RDA}
	\begin{align}
		\green_\mathrm{B}^{(i)} &= \left[ (E+\imag\eta)\mathcal{I} - \hamilton_\mathrm{B}^{(i)} \right]^{-1} \quad\mathrm{with}\quad \hamilton_\mathrm{B}^{(0)} = \hamilton \qquad,\\
		\alpha^{(i+1)} &= \alpha^{(i)}\green^{(i)}_\mathrm{B}\alpha^{(i)} \quad\mathrm{with}\quad \alpha^{(0)} = \coupling \qquad,\\
		\beta^{(i+1)} &= \beta^{(i)}\green^{(i)}_\mathrm{B}\beta^{(i)} \quad\mathrm{with}\quad \beta^{(0)} = \coupling^\dagger \qquad,\\
		\hamilton_\mathrm{L}^{(i+1)} &= \hamilton_\mathrm{L}^{(i)} + \alpha^{(i)}\green_\mathrm{B}^{(i)}\beta^{(i)} \quad\mathrm{with}\quad \hamilton_\mathrm{L}^{(0)} = \hamilton \qquad, \label{JCP2:eqn:RDA:HL}\\
		\hamilton_\mathrm{B}^{(i+1)} &= \hamilton_\mathrm{B}^{(i)} + \alpha^{(i)}\green_\mathrm{B}^{(i)}\beta^{(i)} + \beta^{(i)}\green_\mathrm{B}^{(i)}\alpha^{(i)} \qquad, \label{JCP2:eqn:RDA:HB}\\
		\hamilton_\mathrm{R}^{(i+1)} &= \hamilton_\mathrm{R}^{(i)} + \beta^{(i)}\green_\mathrm{B}^{(i)}\alpha^{(i)} \quad\mathrm{with}\quad \hamilton_\mathrm{R}^{(0)} = \hamilton \qquad. \label{JCP2:eqn:RDA:HR}
	\end{align}%
\end{subequations}%
The effective Hamiltonian matrices \smash{$\hamilton_\mathrm{L/R}^{(i)}$} correspond to the first/last diagonal elements of (\ref{JCP2:eqn:decimation:Aeff}) and get a correction because they are connected to a right/left cell of the bulk.
The effective bulk Hamiltonian matrix \smash{$\hamilton_\mathrm{B}^{(i)}$} gets both corrections because it is connected to both, a left and a right cell.
The effective coupling matrices $\alpha^{(i)}$ and $\beta^{(i)}$ correspond to the upper and lower non-diagonal element of (\ref{JCP2:eqn:decimation:Aeff}).
The RDA is an iterative algorithm, where every iteration step enlarges the system by a factor of 2 until it is large enough to consider the effective Hamiltonian matrices and corresponding Green's functions converged.
\smash{$\hamilton_\mathrm{L/B/R}^{(i)}$} can be assumed to converge if $||\alpha||+||\beta|| \rightarrow 0$.

Finally, the bulk/surface Green's matrix can be calculated as
\begin{equation}
	\green_\mathrm{L/B/R} = \left[ (E+\imag\eta)\mathcal{I} - \hamilton_\mathrm{L/B/R}^{(\infty)}\right]^{-1} \quad.\label{JCP2:eqn:RDA:GLBR}
\end{equation}%

For the $2^k+1$ cells, $k-1$ inversions and $6(k-1)$ multiplications are necessary.
Considering $N$ cells, $\log_2(N-1)-1$ inversions and $6\log_2(N-1)-6$  multiplications have to be done.
This logarithmic scaling of the complexity leads to very fast convergence.

\section{Renormalization-decimation algorithm for electrodes with long unit cells}

The RDA includes the calculation of the Green's function of single UCs, which is a matrix inversion.
Its complexity scales with $(\dim\hamilton)^3$, thus getting worse for larger UCs.
But in cases where the UC is long in transport direction, the short-range interaction leads also to a block-wise tridiagonal UC Hamiltonian matrix in the same way as the block-wise tridiagonal electrode Hamiltonian matrix itself.
Thus, such a UC can be divided into $K$ slices, leading to a $K{\times}K$ shape of the Hamiltonian matrix and the coupling matrix:
\begin{equation}
	\hamilton = \begin{pmatrix}
		\hamilton_1 & \tau_{12} & ~ & ~ \\
		\tau_{21} & \hamilton_2 & \tau_{23} & ~ \\
		~ & \tau_{32} & \mydots & \mydots \\
		~ & ~ & \mydots & \hamilton_K
	\end{pmatrix} \quad,\quad \tau = \begin{pmatrix}
		~ & ~ & ~ & ~ \\
		~ & ~ & ~ & ~ \\
		~ & ~ & ~ & ~ \\
		\tau_{K1} & \phantom{\tau_{12}} & \phantom{\tau_{12}} & \phantom{\tau_{12}}
	\end{pmatrix} \quad,
\end{equation}%
likewise for the total bulk Green's matrix $\green_\text{B}$ of the UC.
Note that -- in contrast to the total electrode Hamiltonian matrix (\ref{JCP2:eqn:Helectrode}) -- this UC Hamiltonian matrix is not periodic.

\subsection{Surface Green's functions}\label{JCP2:sec:DRDA}

\begin{figure}[t!]
	\centering
	\includegraphics{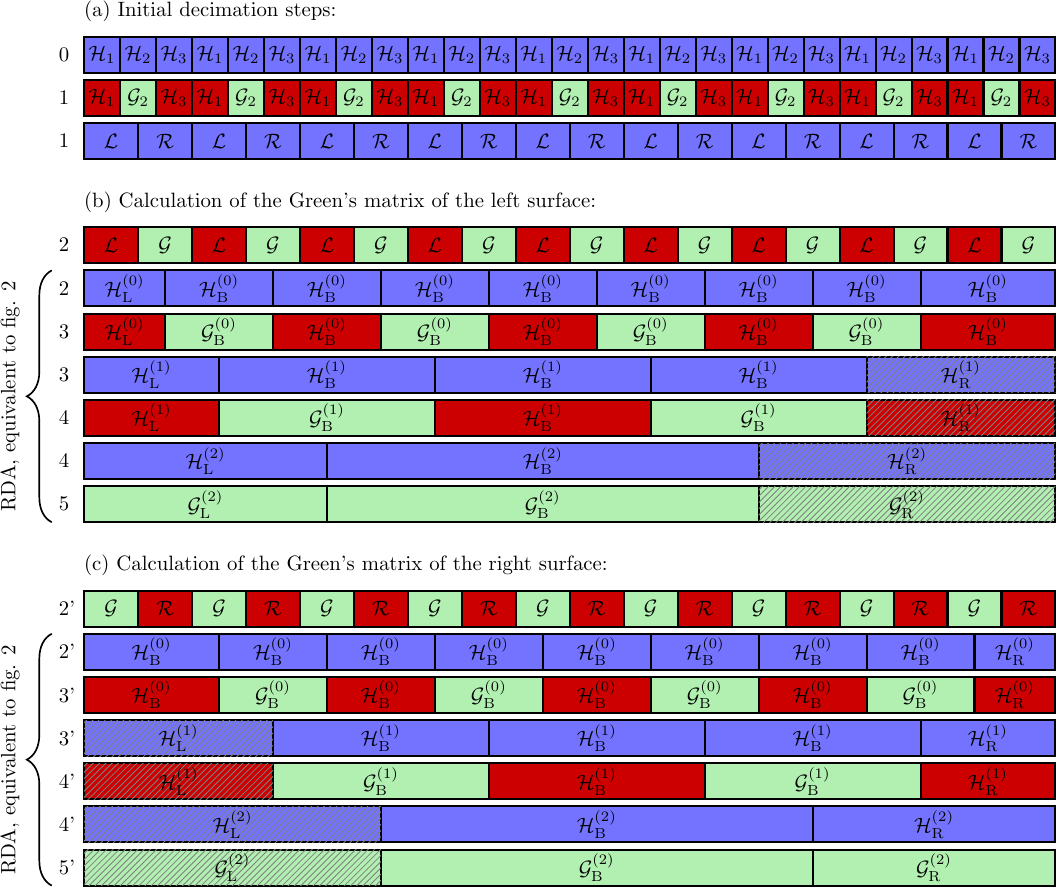}
	\caption[Sketch of the dual RDA]{(Color online.) Sketch of the dRDA for an electrode with 9 UCs, which are each divided into 3 slices $\hamilton_1$, $\hamilton_2$, and $\hamilton_3$. The upper three rows correspond to the initial decimation steps (\ref{JCP2:eqn:DRDA:Decimation}) yielding an effective binary system. The seven middle/lower rows represent the RDA for getting the Green's matrix of the left/right surface. The numbers to the left of each row indicate the iteration step. The color code is the same as in figure~\ref{JCP2:fig:RDA}. Hatched blocks need not be calculated.}\label{JCP2:fig:DualRDA}
\end{figure}

The surface Green's functions $\green_\mathrm{L/R}$ of (\ref{JCP2:eqn:Helectrode}), which are necessary for transport calculations, are now much smaller ones, corresponding to the surface blocks $\hamilton_1$ (for L) and $\hamilton_K$ (for R).
Using the decimation technique one can get rid of the $\hamilton_i$ for $i=2,\ldots,K-1$:
\begin{subequations}
	\label{JCP2:eqn:DRDA:Decimation}
	\begin{align}
		\green_i &= \left[ (E-\imag\eta)\mathcal{I} - \tilde{\hamilton}_i \right]^{-1} \quad\mathrm{with}\quad \tilde{\hamilton}_2 = \hamilton_2 \\
		\tilde{\hamilton}_1^{(i)} &= \tilde{\hamilton}_1^{(i-1)} + \tilde{\coupling}_{1i}\green_i\tilde{\coupling}_{i1} \quad\mathrm{with}\quad \tilde{\hamilton}_1^{(0)} = \hamilton_1 \quad, \label{JCP2:eqn:DRDA:Decimation:H1}\\	
		\tilde{\hamilton}_{i+1} &= \hamilton_{i+1} + \coupling_{(i+1)i}\green_i\coupling_{i(i+1)} \quad,\label{JCP2:eqn:DRDA:Decimation:Hi+1}\\
		\tilde{\coupling}_{1(i+1)} &= \tilde{\coupling}_{1i}\green_i\coupling_{i(i+1)} \quad\mathrm{with}\quad \tilde{\coupling}_{12} = \coupling_{12} \quad,\\
		\tilde{\coupling}_{(i+1)1} &= \coupling_{(i+1)i}\green_i\tilde{\coupling}_{i1} \quad\mathrm{with}\quad \tilde{\coupling}_{21} = \coupling_{21} \quad.
	\end{align}%
\end{subequations}%
Equation~(\ref{JCP2:eqn:DRDA:Decimation:H1}) means that the cell to the left of the decimated one, which is always the first one, is modified.
Equation~(\ref{JCP2:eqn:DRDA:Decimation:Hi+1}) means that the cell to the right of the decimated one is modified.
This results in an effective periodic binary system
\begin{equation}
	\hamilton_\mathrm{el}^\mathrm{eff} = \begin{pmatrix}
		\mathcal{L} & \sigma_\mathcal{LR} & ~ & ~ & ~ & ~ & ~ \\
		\sigma_\mathcal{RL} & \mathcal{R} & \tau_\mathcal{RL} & ~ & ~ & ~ & ~ \\
		~ & \tau_\mathcal{LR} & \mathcal{L} & \sigma_\mathcal{LR} & ~ & ~ & ~ \\
		~ & ~ & \sigma_\mathcal{RL} & \mathcal{R} & \tau_\mathcal{RL} & ~ & ~ \\
		~ & ~ & ~ & \tau_\mathcal{LR} & \mydots & \mydots & ~ \\
		~ & ~ & ~ & ~ & \mydots & \mathcal{L} & \sigma_\mathcal{LR} \\
		~ & ~ & ~ & ~ & ~ & \sigma_\mathcal{RL} & \mathcal{R}
	\end{pmatrix} \label{JCP2:eqn:DRDA:H_el^eff}
\end{equation}%
with effective left matrices $\mathcal{L} = \tilde{\hamilton}_1^{(K-1)}$, effective right matrices $\mathcal{R} = \tilde{\hamilton}_K$, and effective coupling matrices $\sigma_\mathcal{LR} = \tilde{\coupling}_{1K}$, $\sigma_\mathcal{RL} = \tilde{\coupling}_{K1}$, $\tau_\mathcal{RL} = \coupling_{K1}$, and $\tau_\mathcal{LR} = \coupling_{1K}$, as visualized in figure \ref{JCP2:fig:DualRDA}(a).

The surface Green's matrices of the effective binary system can be evaluated by doing two separate RDA calculations.
A graphical visualization is shown in figure \ref{JCP2:fig:DualRDA}(b,c).
First, the left surface $\green_\mathrm{L}$ can be calculated in the following way:
The decimation of all $\mathcal{R}$ in (\ref{JCP2:eqn:DRDA:H_el^eff}) leads to a new effective periodic Hamiltonian matrix, except for the lower right block (in which we are not interested at this time).
Using the results as initial values for the application of the RDA (\ref{JCP2:eqn:RDA}) yields
\begin{subequations}
	\label{JCP2:eqn:DRDA:L}
	\begin{align}
		\green_\mathcal{R} &= \left[ (E+\imag\eta)\mathcal{I} - \mathcal{R}\right]^{-1} \quad,\\
		\hamilton_{\mathrm{L}}^{(0)} &= \mathcal{L} + \sigma_\mathcal{LR}\green_\mathcal{R}\sigma_\mathcal{RL} \quad,\\
		\hamilton_{\mathrm{B}}^{(0)} &= \mathcal{L} + \sigma_\mathcal{LR}\green_\mathcal{R}\sigma_\mathcal{RL} + \tau_\mathcal{LR}\green_\mathcal{R}\tau_\mathcal{RL} \quad, \label{JCP2:eqn:DRDA:L:B}\\
		\alpha^{(0)} &= \sigma_\mathcal{LR}\green_\mathcal{R}\tau_\mathcal{RL} \quad,\\
		\beta^{(0)} &= \tau_\mathcal{LR}\green_\mathcal{R}\sigma_\mathcal{RL} \quad .
	\end{align}%
\end{subequations}%
The execution of (\ref{JCP2:eqn:RDA:HR}) is not necessary.
Afterwards, $\green_\mathrm{L}$ is determined by (\ref{JCP2:eqn:RDA:GLBR}).
Second, the right surface $\green_\mathrm{R}$ can be calculated in an equivalent way:
The decimation of all $\mathcal{L}$ in (\ref{JCP2:eqn:DRDA:H_el^eff}) leads to a new effective periodic Hamiltonian matrix, except for the upper left block (in which we are also not interested at this time).
Using the results as initial values for the application of the RDA (\ref{JCP2:eqn:RDA}) a second time yields
\begin{subequations}
	\label{JCP2:eqn:DRDA:R}
	\begin{align}
		\green_\mathcal{L} &= \left[ (E+\imag\eta)\mathcal{I} - \mathcal{L}\right]^{-1} \quad,\\
		\hamilton_{\mathrm{R}}^{(0)} &= \mathcal{R} + \sigma_\mathcal{RL}\green_\mathcal{L}\sigma_\mathcal{LR} \quad,\\
		\hamilton_{\mathrm{B}}^{(0)} &= \mathcal{R} + \sigma_\mathcal{RL}\green_\mathcal{L}\sigma_\mathcal{LR} + \tau_\mathcal{RL}\green_\mathcal{L}\tau_\mathcal{LR} \quad, \label{JCP2:eqn:DRDA:R:B}\\
		\alpha^{(0)} &= \tau_\mathcal{RL}\green_\mathcal{L}\sigma_\mathcal{LR} \quad,\\
		\beta^{(0)} &= \sigma_\mathcal{RL}\green_\mathcal{L}\tau_\mathcal{LR} \quad.
	\end{align}%
\end{subequations}%
Here, the execution of (\ref{JCP2:eqn:RDA:HL}) is not necessary.
Afterwards, $\green_\mathrm{R}$ is determined by (\ref{JCP2:eqn:RDA:GLBR}).
Because the RDA has to be done twice, we call the total algorithm dual RDA (dRDA).

\subsection{Bulk Green's functions and electrode density of states}\label{JCP2:sec:MRDA}

The total periodic Green's function with an infinite number of long UCs each consisting of $K$ slices reads
\begin{equation}
	\green_\text{el} =  \begin{pmatrix}
		\mydots & \mydots & \mydots & \mydots & \mydots & \mydots & \mydots & \mydots & \mydots & \mydots \\
		\mydots & \green_{11} & \green_{12} & \mydots & \mydots & \mydots & \mydots & \mydots & \mydots & \mydots \\
		\mydots & \green_{21} & \green_{22} & \mydots & \mydots & \mydots & \mydots & \mydots & \mydots & \mydots \\
		\mydots & \mydots & \mydots & \mydots & \green_{(K-1)K} & \mydots & \mydots & \mydots & \mydots & \mydots \\
		\mydots & \mydots & \mydots & \green_{K(K-1)} & \green_{KK} & \green_{K1} & \mydots & \mydots & \mydots & \mydots \\
		\mydots & \mydots & \mydots & \mydots & \green_{1K} & \green_{11} & \green_{12} & \mydots & \mydots & \mydots \\
		\mydots & \mydots & \mydots & \mydots & \mydots & \green_{21} & \green_{22} & \mydots & \mydots & \mydots \\
		\mydots & \mydots & \mydots & \mydots & \mydots & \mydots & \mydots & \mydots & \green_{(K-1)K} & \mydots \\
		\mydots & \mydots & \mydots & \mydots & \mydots & \mydots & \mydots & \green_{K(K-1)} & \green_{KK} & \mydots \\
		\mydots & \mydots & \mydots & \mydots & \mydots & \mydots & \mydots & \mydots & \mydots & \mydots \\
	\end{pmatrix} \quad.
\end{equation}%
The density of states of each electrode as a representative for its electronic structure is calculated by
\begin{equation}
	\dos_\mathrm{el}(E) = -\frac{1}{\pi}\Im\,\Tr\left(\green_\mathrm{el}\overlap_\mathrm{el}\right)_\mathrm{B} = -\frac{1}{\pi}\sum_{i=1}^K\Im\left[ \Tr\left(\green_{ii}\overlap_{ii}\right) + \Tr\left(\green_{ij}\overlap_{ji}\right) + \Tr\left(\green_{ji}\overlap_{ij}\right) \right] \quad,
\end{equation}%
where $j = (i+1)\,\text{mod}\,K$.
Considering cell $i$ as the rightmost slice and cell $j$ as the leftmost slice, the slices $j+1$,\ldots,$i-1$ have to be decimated, similar to (\ref{JCP2:eqn:DRDA:Decimation}).
This leads to the following effective overall bulk Hamiltonian matrix and Green's function:
\begin{equation}
	\begin{pmatrix}
		\mydots & \mydots & ~ & ~ & ~ & ~ & \\
		\mydots & E-\mathcal{L}_j & -\sigma_{ji} & ~ & ~ & ~ & \\
		~ & -\sigma_{ij} & E-\mathcal{R}_i & -\tau_{ij} & ~ & ~ & \\
		~ & ~ & -\tau_{ji} & E-\mathcal{L}_j & -\sigma_{ji} & ~ & \\
		~ & ~ & ~ & -\sigma_{ij} & E-\mathcal{R}_i & \mydots & \\
		~ & ~ & ~ & ~ & \mydots & \mydots &
	\end{pmatrix} \begin{pmatrix}
		\mydots & \mydots & \mydots & \mydots & \mydots & \mydots \\
		\mydots & \green_{jj} & \mydots & \mydots & \mydots & \mydots \\
		\mydots & \mydots & \green_{ii} & \green_{ij} & \mydots & \mydots \\
		\mydots & \mydots & \green_{ji} & \green_{jj} & \mydots & \mydots \\
		\mydots & \mydots & \mydots & \mydots & \green_{ii} & \mydots \\
		\mydots & \mydots & \mydots & \mydots & \mydots & \mydots \\
	\end{pmatrix} = \mathcal{I}
\end{equation}%
Focusing on the center 2$\times$2 matrix block, decimating the $\mathcal{L}_j$ in the left half-infinite part, decimating the $\mathcal{R}_i$ in the right half-infinite part, and applying the RDA to both parts yields
\begin{equation}
	\begin{pmatrix}
		\mydots & \mydots & ~ & ~ & ~ & ~ & \\
		\mydots & E-\hamilton_{i,\mathrm{B}}^{(\infty)} & 0 & ~ & ~ & ~ & \\
		~ & 0 & E-\hamilton_{i,\mathrm{R}}^{(\infty)} & -\tau_{ij} & ~ & ~ & \\
		~ & ~ & -\tau_{ji} & E-\hamilton_{j,\mathrm{L}}^{(\infty)} & 0 & ~ & \\
		~ & ~ & ~ & 0 & E-\hamilton_{j,\mathrm{B}}^{(\infty)} & \mydots & \\
		~ & ~ & ~ & ~ & \mydots & \mydots &
	\end{pmatrix} \begin{pmatrix}
		\mydots & \mydots & \mydots & \mydots \\
		\mydots & \green_{ii} & \green_{ij} & \mydots \\
		\mydots & \green_{ji} & \green_{jj} & \mydots \\
		\mydots & \mydots & \mydots & \mydots \\
	\end{pmatrix} =\mathcal{I} \quad.
\end{equation}%
Now, the center 2$\times$2 matrix block can be solved in an isolated way to get the necessary diagonal elements and first upper/lower non-diagonal elements
\begin{subequations}
	\label{JCP2:eqn:DRDA:greenij}
	\begin{align}
		\green_{ii} &= \left[ (E-\imag\eta)\mathcal{I} - \hamilton_{i,\mathrm{R}}^{(\infty)} - \tau_{ij}\green_{j,\mathrm{L}}\tau_{ji} \right]^{-1}\quad,\\
		\green_{jj} &= \left[ (E-\imag\eta)\mathcal{I} - \hamilton_{j,\mathrm{L}}^{(\infty)} - \tau_{ji}\green_{i,\mathrm{R}}\tau_{ij} \right]^{-1}\quad,\\
		\green_{ij} &= \green_{i,\mathrm{R}}\tau_{ij}\green_{jj} \quad,\\
		\green_{ji} &= \green_{j,\mathrm{L}}\tau_{ji}\green_{ii} \quad.
	\end{align}%
\end{subequations}%
In summary, for each $1\leq i\leq K$, decimations similar to (\ref{JCP2:eqn:DRDA:Decimation}), the RDA, and the final calculations (\ref{JCP2:eqn:DRDA:greenij}) have to be done to get all the values to calculate the electrode density of states.
Because the RDA has to be done $K$ times, we call the total algorithm multiple RDA (mRDA).
Note that the quantities which are necessary for electron transport, namely the surface Green's functions, are included:
\begin{subequations}
	\begin{align}
		\green_\mathrm{L} &= \left[ (E-\imag\eta)\mathcal{I} - \hamilton_{1,\mathrm{L}}^{(\infty)} \right]^{-1} \quad,\\
		\green_\mathrm{R} &= \left[ (E-\imag\eta)\mathcal{I} - \hamilton_{K,\mathrm{R}}^{(\infty)} \right]^{-1} \quad.
	\end{align}%
\end{subequations}%

\section{Complexity measure and performance test}

\begin{figure}[!t]
	\includegraphics{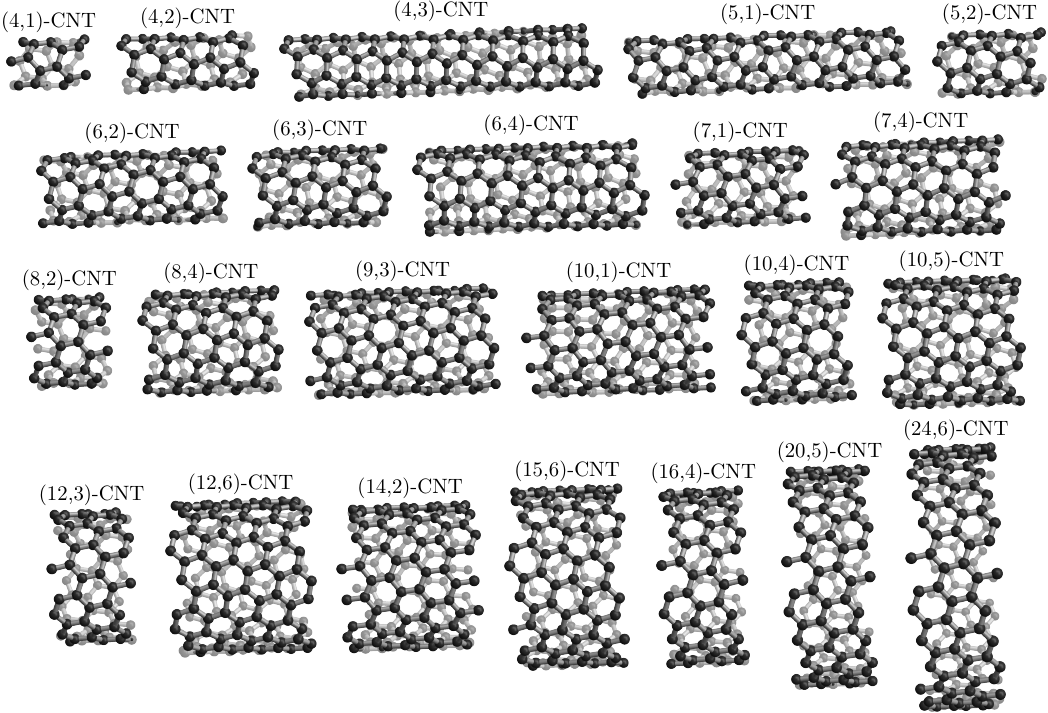}
	\caption[Geometric structures of the CNT unit cells]{CNT UCs to which the different versions of the RDA are applied for the performance test.}\label{JCP2:fig:CNTs}
\end{figure}

In the following, the dRDA and the mRDA are compared to the conventional RDA (cRDA) of section \ref{JCP2:sec:RDA} by calculating the surface and bulk Green's functions for 1000 energies.
For this, we look at complexity measures representing the calculation time $t$ and the memory requirement (RAM) $m$ of a self-written C++ implementation using LAPACK routines~\cite{LAPACK}.
We apply the algorithms to the chiral CNTs shown in figure \ref{JCP2:fig:CNTs}.
Different diameters and different chiralities, which strongly affect the UC length, are addressed.
The number of subdivisions differs from $K=2$ to $K=9$.
$K$ and $\dim\hamilton$ for each CNT are listed in table \ref{JCP2:tab:CNTs}.
Because the chiral CNTs are semiconducting, their bandgap depends on the diameter and the chirality.
In energy regions within the bandgap with nearly no transmission, the RDA converges faster than in other energy regions.
Therefore, we focus on 1000 energies outside the bandgap to get comparable calculation times for different CNTs.

\begin{figure}[t]
	\includegraphics{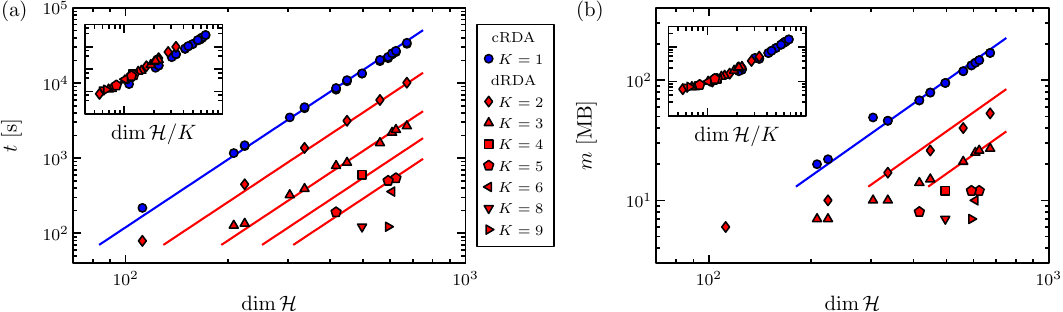}
	\caption[Calculation time and memory requirement for the dual RDA]{(Color online.) (a) Calculation time $t$ and (b) memory requirement (RAM) $m$ for the cRDA (blue) and the dRDA for long UCs (red) when only calculating the surface Green's functions. The algorithms are applied to the different CNTs shown in figure \ref{JCP2:fig:CNTs}. The solid lines correspond to the complexity measures (\ref{JCP2:eqn:complexity:RDA:t}--\ref{JCP2:eqn:complexity:DRDA:m}). The insets show the same data, but the x~axes are scaled with the number of slices $K$ ($K=1$ for the cRDA, $K=2\ldots 7$ for the dRDA).}\label{JCP2:fig:performance:DRDA}
\end{figure}

Figure \ref{JCP2:fig:performance:DRDA} shows $t$ and $m$ for the dRDA compared to the cRDA.
First of all, it can be seen that for all chiral CNTs $t$ and $m$ are reduced drastically by at least a factor of 3 for the shortest CNT UCs up to a factor of 176 for the longest CNT UCs studied here.
Longer UCs lead to even larger reductions.
The dependencies can be well described as follows.
The calculation time is characterized by the number of arithmetic operations.
The cRDA~(\ref{JCP2:eqn:RDA}) needs 1~matrix inversion, 6~matrix multiplications, and 4~matrix additions in each iteration step and $a$~iterations in total.
For the CNTs we usually get convergence after approximately 15 to 25 iterations.
The dominant operations are the matrix inversions, which scale approximately with the third power of the dimension.
This leads to
\begin{equation}
	t_\text{cRDA} \sim a\left(\dim\hamilton\right)^3 \quad.\label{JCP2:eqn:complexity:RDA:t}
\end{equation}%
The RAM is characterized by the number stored matrix entries.
During the cRDA, 5 matrices ($\hamilton_\text{L/B/R}$, $\alpha$, and $\beta$), which are updated in each iteration step, are stored.
3 additional cRDA matrices are stored to omit repeating the same matrix multiplications.
This leads to
\begin{equation}
	m_\text{cRDA} \sim 8\left(\dim\hamilton\right)^2 \quad.\label{JCP2:eqn:complexity:RDA:m}
\end{equation}%
Both complexity measures are depicted in figure \ref{JCP2:fig:performance:DRDA} and show the linear dependence in the doubly logarithmic scale (blue lines).
An increase of 1~order of magnitude of $\dim\hamilton$ leads to an increase of 3~orders of magnitude in~$t$ and 2~orders of magnitude in~$m$.

\begin{table}[!t]
	\begin{tabular}{c|c|c}
		$K$ & CNT & $\text{dim}\,\hamilton$ \\
		\hline
		2 & (4,1)  & 112 \\
		2 & (8,2)  & 224 \\
		2 & (12,3) & 336 \\
		2 & (16,4) & 448 \\
		2 & (20,5) & 560 \\
		2 & (24,6) & 672
	\end{tabular}\hspace{1em}
	\begin{tabular}{c|c|c}
		$K$ & CNT & $\text{dim}\,\hamilton$ \\
		\hline
		3 & (4,2)  & 224 \\
		3 & (5,2)  & 208 \\
		3 & (6,3)  & 336 \\
		3 & (7,1)  & 304 \\
		3 & (8,4)  & 448 \\
		3 & (10,4) & 416
	\end{tabular}\hspace{1em}
	\begin{tabular}{c|c|c}
		$K$ & CNT & $\text{dim}\,\hamilton$ \\
		\hline
		3 & (10,5) & 560 \\
		3 & (12,6) & 672 \\
		3 & (14,2) & 608 \\
		3 & (15,6) & 624 \\
		4 & (7,4)  & 496 \\
		5 & (6,2)  & 416
	\end{tabular}\hspace{1em}
	\begin{tabular}{c|c|c}
		$K$ & CNT & $\text{dim}\,\hamilton$ \\
		\hline
		5 & (9,3)  & 624 \\
		5 & (10,1) & 592 \\
		6 & (6,4)  & 608 \\
		8 & (5,1)  & 496 \\
		9 & (4,3)  & 592 \\
		\multicolumn{3}{c}{~}
	\end{tabular}
	\caption[Number of slices and dimension of the Hamiltonian matrix for different CNTs]{Number of slices $K$ and dimension of the UC Hamiltonian matrix for the CNTs in figure \ref{JCP2:fig:CNTs}.}\label{JCP2:tab:CNTs}
\end{table}

The complexity measures of the dRDA described in section \ref{JCP2:sec:MRDA} can be similarly obtained.
The dRDA needs $K-1$~decimation steps~(\ref{JCP2:eqn:DRDA:Decimation}), (\ref{JCP2:eqn:DRDA:L}), and (\ref{JCP2:eqn:DRDA:R}) and $2$~RDA executions~(\ref{JCP2:eqn:RDA}).
Each decimation step needs 1~matrix inversion, 6~matrix multiplications, 2~matrix additions.
In total this yields $K+2a$ matrix inversions, $6K+12a$ matrix multiplications, and $2K+8a$ matrix additions.
In comparison to the cRDA, here the dimension of the matrices is much smaller.
To get the fastest algorithm, the $K$~slices should be as short as possible and thus the corresponding matrices have similar dimension.
So, we can assume that the dimension of each matrix is smaller than $\dim\hamilton$ by a factor of~$K$.
This leads to
\begin{subequations}
	\label{JCP2:eqn:complexity:DRDA:t}
	\begin{empheq}[left={t_\text{dRDA} \sim (K+2a)\left(\dfrac{\dim\hamilton}{K}\right)^3 \sim \empheqlbrace}]{align}
		2a\left(\dfrac{\dim\hamilton}{K}\right)^3 \quad &\text{for}~K \ll a \quad,\label{JCP2:eqn:complexity:DRDA:t:A}\\
		K\left(\dfrac{\dim\hamilton}{K}\right)^3 \quad &\text{for}~K \gg a \quad.\label{JCP2:eqn:complexity:DRDA:t:B}
	\end{empheq}%
\end{subequations}%
\newpage
\noindent During the dRDA, 4 additional matrices ($\mathcal{L}$, $\mathcal{R}$, $\sigma_\mathcal{LR}$ and $\sigma_\mathcal{RL}$) are stored, leading to
\begin{equation}
	m_\text{dRDA} \sim 12\left(\dfrac{\dim\hamilton}{K}\right)^2 \quad.\label{JCP2:eqn:complexity:DRDA:m}
\end{equation}%
The data in figure \ref{JCP2:fig:performance:DRDA} (red) confirm these dependencies.
For all CNTs with similar length and thus, similar number of slices $K$, the same linear dependence can be seen in the doubly logarithmic scale (indicated by four lines), but reduced by a constant factor.
In the case of the calculation time $t$, this reduction equals $K^3/2$, which is the ratio of (\ref{JCP2:eqn:complexity:RDA:t}) and~(\ref{JCP2:eqn:complexity:DRDA:t:A}).
The lines in figure \ref{JCP2:fig:performance:DRDA} fulfill this relation with $a=12$.
The insets of figure \ref{JCP2:fig:performance:DRDA} show the same data, but with an abscissa, which is scaled with $1/K$.
Here, all data points of the dRDA fall onto one curve corresponding to (\ref{JCP2:eqn:complexity:DRDA:t}) and (\ref{JCP2:eqn:complexity:DRDA:m}) for the case $K\ll a$.
The difference of the calculation time between the cRDA and the dRDA in the inset of figure \ref{JCP2:fig:performance:DRDA}(a) is the factor of 2 caused by the 2 separate RDA calculations during the dRDA.

\begin{figure}[b]
	\centering
	\includegraphics{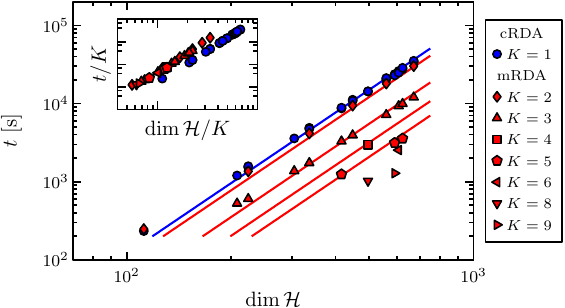}
	\caption[Calculation time for the multiple RDA]{(Color online.) Same as figure \ref{JCP2:fig:performance:DRDA}(a) but for the mRDA, where the bulk Green's function is additionally computed. The solid lines correspond to the complexity measures (\ref{JCP2:eqn:complexity:RDA:t}) and (\ref{JCP2:eqn:complexity:MRDA:t}).}\label{JCP2:fig:performance:MRDA}
\end{figure}

For the mRDA, where the bulk Green's function for the electrode density of states is additionally calculated, the comparison can be done in a similar way.
The calculation times compared to the cRDA are shown in figure \ref{JCP2:fig:performance:MRDA}.
For the cRDA nothing changes, because the calculation of the total bulk Green's function is necessary for the calculation of the surface Green's functions.
This leads to the same complexity measures (\ref{JCP2:eqn:complexity:RDA:t}) and (\ref{JCP2:eqn:complexity:RDA:m}).
For the mRDA/dRDA this is not the case, because the calculation of the total bulk Green's function (mRDA) can be partly omitted when one wants to obtain the surface Green's functions (dRDA) only.
Instead of calculating two surface blocks during the dRDA, all $K$ bulk blocks have to be computed during the mRDA.
In total, for each of these $K$ blocks the mRDA described in section \ref{JCP2:sec:MRDA} needs $K-1$ decimation steps (1 matrix inversion, 6 matrix multiplications, and 2 matrix additions each), followed by $2$ RDA executions, and afterwards additional 4 matrix inversions, 4 matrix multiplications, and 2 matrix additions to obtain the necessary diagonal and first non-diagonal matrix blocks.
In total this yields $K(K+3+2a)$ matrix inversions, $K(6K-2+12a)$ matrix multiplications, and $K(2K+8a)$ matrix additions.
As the dimension of the matrices is a factor of $K$ smaller, the complexity measure yields
\begin{subequations}
	\label{JCP2:eqn:complexity:MRDA:t}
	\begin{empheq}[left={t_\text{mRDA} \sim K(K+3+2a)\left(\dfrac{\dim\hamilton}{K}\right)^3 \sim \empheqlbrace}]{align}
		2aK\left(\dfrac{\dim\hamilton}{K}\right)^3 \quad &\text{for}~K \ll a \quad,\label{JCP2:eqn:complexity:MRDA:t:A}\\
		K^2\left(\dfrac{\dim\hamilton}{K}\right)^3 \quad &\text{for}~K \gg a \quad.\label{JCP2:eqn:complexity:MRDA:t:B}
	\end{empheq}%
\end{subequations}%
The data are shown in figure \ref{JCP2:fig:performance:MRDA} and confirm these dependencies.
In the same way as for the dRDA, a linear dependence is valid in the doubly logarithmic scale for CNTs with the same number of slices $K$.
In the inset the abscissa and the ordinate are scaled with $1/K$ to verify (\ref{JCP2:eqn:complexity:MRDA:t:A}).
As all red points lie on the same curve, this dependence is valid.
The additional factor $K$ compared to the dRDA describes the $K$fold amount of information which must be computed for the bulk Green's function compared to the surface Green's functions.
Thus, the calculation time reduction is lower compared to the dRDA.
There is a small improvement for the shortest CNT UCs and a reduction of computation time up to a factor of 18 for the longest CNT UCs studied here.
The difference between the cRDA and the mRDA is the factor of 2 caused by the 2 RDA calculations which are done for each bulk matrix block.

\section{Exemplary results}

\begin{figure}[t!]
	\includegraphics{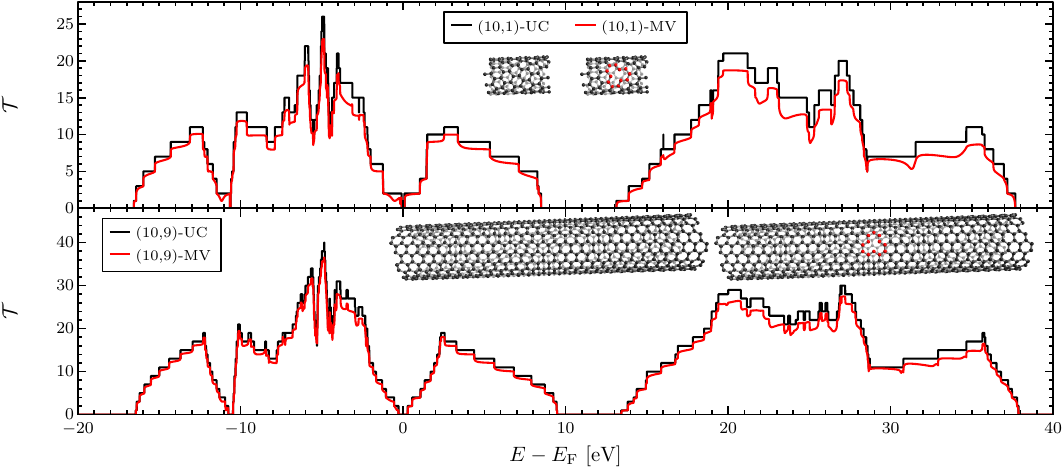}\\[0.5em]
	\includegraphics{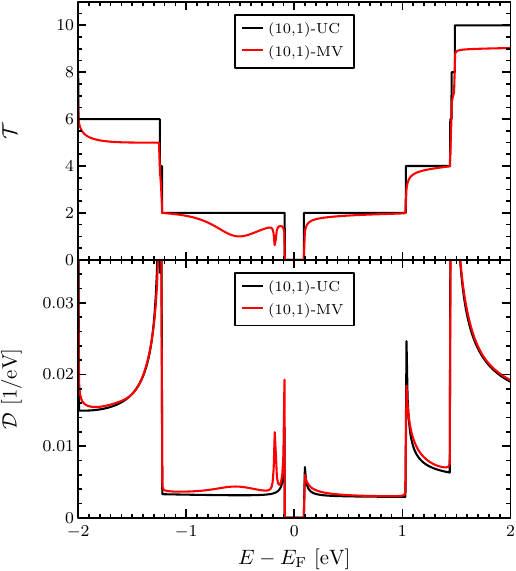}\hfill
	\includegraphics{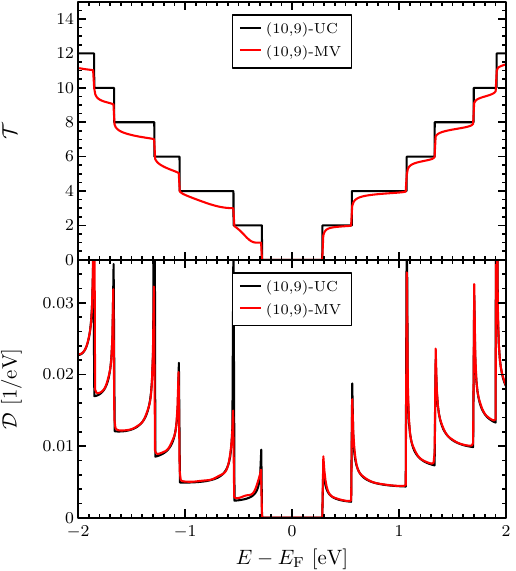}
	\caption[Transmission spectrum and device density of states for CNTs]{(Color online.) (upper) Complete transmission spectrum $\mathcal{T}$ of the (10,1)-CNT and the (10,9)-CNT for the ideal, periodic case (black) and for a CNT with one MV (red). (center) Same results around the Fermi energy. (lower) Device density of states $\mathcal{D}$ for the same structures.}\label{JCP2:fig:T(E)}
\end{figure}

\begin{figure}
	\includegraphics{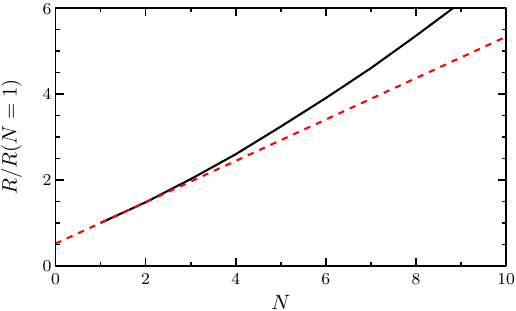}\hfill
	\includegraphics{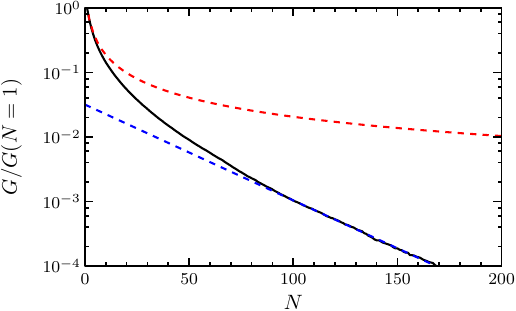}
	\caption[Resistance and conductance for defective CNTs]{(Color online.) Resistance $R$ and conductance $G$ of the (10,1)-CNT as a function of the number of MV defects (black solid lines) at $300\,\text{K}$. The data are normalized to the values for one MV defect. The dashed lines are regressions concerning the diffusion regime (red) and the localization regime (blue).}\label{JCP2:fig:G(N)}
\end{figure}

The aim of the RDA and our improved versions is the usage for quantum transport calculations.
We present some exemplary results for CNTs to show, that it is possible to calculate transport properties of materials with huge UCs very efficiently.
We use a DFTB model~\cite{PhysRevB.51.12947,IntJQuantumChem.58.185} and the parameter set 3ob of Gaus et al.~\cite{JChemTheoryComput.9.338}, which is suitable for carbon materials and includes the 2s and the three 2p Slater-Koster orbitals~\cite{PhysRev.94.1498}.
We consider two CNTs: (A)~the (10,1)-CNT as one of the biggest shown in figure \ref{JCP2:fig:CNTs}, containing 148 atoms within one UC, and (B) the even bigger (10,9)-CNT, containing 1084 atoms within one UC.
Figure \ref{JCP2:fig:T(E)} presents the transmission spectra for the periodic (10,1)- and (10,9)-CNTs, the transmission spectra of the CNTs which include a monovacancy (MV -- one removed atom with relaxation of the neighbouring atomic structure), the electrode density of states of the periodic CNTs, and the device density of states of the CNTs with one MV.
For both CNTs the MV defect leads to a significant reduction of the transmission, especially at energies slightly below the Fermi energy.
For the (10,1)-CNT two features can be seen: a broad dip at approximately $0.5\,\mathrm{eV}$ below $E_\mathrm{F}$ and a sharp dip at approximately $0.2\,\mathrm{eV}$ below $E_\mathrm{F}$.
Comparing this with the device density of states, it can be seen that the dips in the transmission coincide with similar broad/sharp peaks in the density of states.
These peaks are defect states which cause a localization of the wave function at the defect and thus reduce the transmission significantly.
The (10,9)-CNT with a MV shows no significant features in the density of states and the transmission spectrum around the Fermi energy.

Figure \ref{JCP2:fig:G(N)} shows the dependence of the resistance $R$ and the conductance $G$ at $300\,\text{K}$ on the number of monovacancies $N$ within a (10,1)-CNT, which is $8518\,\text{\AA}$ long, i.e. consists of 569 UCs.
The calculations have been performed using the presented methods and the RGF to treat huge central regions~\cite{JPhysCSolidStatePhys.14.235}.
The data can be described by two distinct transport regimes, the diffusion regime~\cite{RevModPhys.69.731} and the strong localization regime~\cite{PhysRevLett.47.1546,PhysRevLett.42.673}.
Regressions concerning the corresponding regimes in figure \ref{JCP2:fig:G(N)} fit the data very well.
We get $R=0.52\,(1+N/1.1)$ for the diffusion regime and $G=0.032\,\text{exp}(-N/29)$ for the localization regime, giving the dimensionless elastic mean free path (normalized to the average defect distance) $N^\text{mfp}=1.1$ and the dimensionless localization length $N^\text{loc}=29$.
Further results of electron transport through arbitrary chiral CNTs can be found in our comprehensive study~\cite{JPhysCommun.2.105012}.

The calculation of the surface Green's functions (not containing the treatment of the central region) for the (10,1)-CNT using the dRDA is a factor 45 faster compared to using the cRDA due to a subdivision into five slices.
The (10,9)-CNT is not treatable with the cRDA within acceptable time as it would take roughly 400 times longer than for the (10,1)-CNT.
Thus, the calculation time reduction for the (10,9)-CNT, which is subdivided into 24 slices, can be estimated to be about 4 orders of magnitudes for the dRDA compared to the cRDA.
For quantum transport calculations of mesoscopic defective CNTs in the $\upmu$m-range the RGF usually dominates the calculation time for short electrode UCs.
For larger cells like the (10,1)-CNT the electrode calculation time using the cRDA becomes of the order of the RGF calculation time for the central region.
For this case, the usage of the dRDA/mRDA reduces the electrode calculation time back to a negligible value, making its application very useful.
Furthermore, the dRDA/mRDA enables the electrode calculation and thus also the computation of transport through mesoscopic defective systems for even longer UCs like for the (10,9)-CNT.

\section{Summary and conclusions}

In the first part, we presented an improved electrode algorithm for tight-binding-based quantum transport calculations of systems with very long UCs in transport direction, like it is the case, e.g., for chiral nanotubes and chiral nanoribbons, or for nanotubes with periodic defects.
The Hamiltonian matrix of the electrodes is block-wise tridiagonal and always periodic, what is used in the RDA to calculate the bulk Green's function and the left/right surface Green's functions.
For the case of very long UCs each matrix block is again block-wise tridiagonal, but not periodic.
We showed how the RDA can be improved for such cases by dividing the UCs into $K$~slices.
We derived iterative equations for the surface Green's functions, which are needed to calculate the transmission function, as well as for the diagonal and the first non-diagonal blocks of the bulk Green's functions, necessary to calculate the bulk density of states.

In the second part, we analyzed the complexity of the improved RDA.
We showed that a factor $K^3$ is gained for the transmission computation.
If the bulk density of states is calculated, a factor $K^2$ is gained.
We applied the dRDA/mRDA to calculate the surface and bulk Green's function for various CNTs of different UC length.
The extracted computation times verify the complexity scaling we predicted.
The presented algorithm allows us to calculate electron transport through quasi one-dimensional systems with long UCs much more efficiently than the cRDA, as exemplarily discussed for transport through (10,1)- and (10,9)-CNTs with MV defects.
The dRDA is the key for treating extremely long UCs like the (10,9)-CNT, which would not be possible with the cRDA.

This work contributes to the continuing development of numerical implementations in quantum transport theory.
It may be useful for reducing the computation time of electron transport calculations of different systems with long UCs like nanotubes, nanoribbons, and nanowires, as well as for defective supercell calculations.

\section*{Acknowledgement}

This work is funded by the European Union (ERDF) and the Free State of Saxony via the ESF project 100231947 (Young Investigators Group Computer Simulations for Materials Design - CoSiMa).
\begin{center}
	\includegraphics[width=0.4\textwidth]{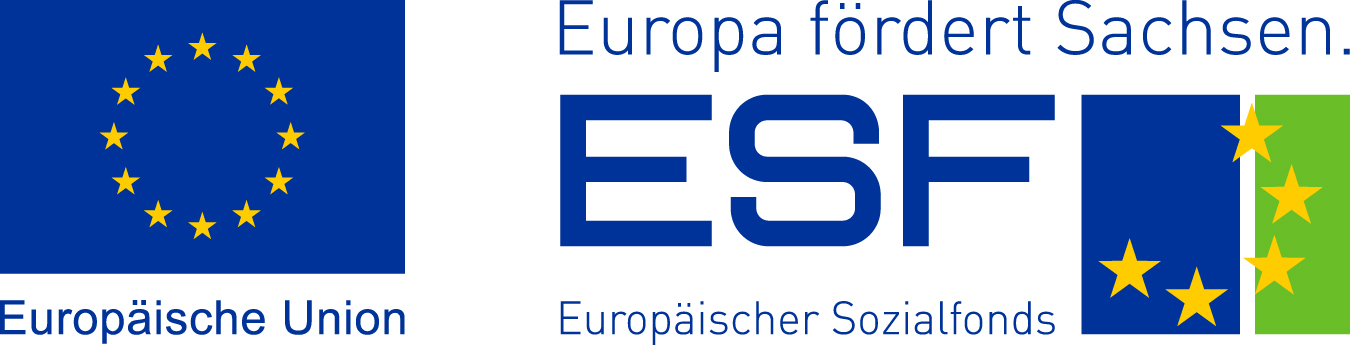}
\end{center}

\end{document}